\documentclass[12pt]{article}

\usepackage{epsfig}
\usepackage{latexsym}
\usepackage{axodraw}
\usepackage{amssymb}

\begin{document}

\begin{titlepage}

\baselineskip 24pt

\begin{center}

{\Large {\bf Kinematical Lepton Transmutation in $e^+ e^-$ Collision 
and Vector Boson Decay}}

\vspace{.5cm}

\baselineskip 14pt

{\large Jos\'e BORDES}\\
jose.m.bordes\,@\,uv.es\\
{\it Departament Fisica Teorica, Universitat de Valencia,\\
  calle Dr. Moliner 50, E-46100 Burjassot (Valencia), Spain}\\
\vspace{.2cm}
{\large CHAN Hong-Mo}\\
chanhm\,@\,v2.rl.ac.uk \\
{\it Rutherford Appleton Laboratory,\\
  Chilton, Didcot, Oxon, OX11 0QX, United Kingdom}\\
\vspace{.2cm}
{\large TSOU Sheung Tsun}\\
tsou\,@\,maths.ox.ac.uk\\
{\it Mathematical Institute, University of Oxford,\\
  24-29 St. Giles', Oxford, OX1 3LB, United Kingdom}

\end{center}

\vspace{.3cm}

\begin{abstract}

The change in orientation in generation space (rotation) of the fermion 
mass matrix with changing scales can lead to flavour-violations through
just the kinematics of a non-diagonal mass matrix.  Such effects for the
reactions: $e^+ e^- \longrightarrow e^\pm \mu^\mp, e^\pm \tau^\mp, 
\mu^\pm \tau^\mp$, and for the decays of vector bosons into the same 
channels, are calculated following a method suggested earlier which 
gives the differential cross section for each reaction and the branching 
ratio for each decay mode in terms of an overall normalization depending 
only on the speed at which the mass matrix rotates.  A rotation speed
estimated earlier, under certain assumptions from the fermion mixing angles 
and mass ratios, is found to give the above effects at a level readily
detectable in modern high sensitivity experiments such as Bepc, Cleo, 
BaBar and Belle, at least in principle.  The observation of these effects 
would not only confirm the concept of a rotating mass matrix with a 
significance on par with the running coupling constant, but also offer 
some valuable insight into the origin of fermion generations.  However, 
a negative result cannot unfortunately rule out the rotating mass 
matrix since the effects deduced here from this mechanism
alone could in principle 
be cancelled by other rotation effects.   

\end{abstract}

\end{titlepage}

\clearpage

\baselineskip 14pt

\setcounter{equation}{0}

\section{Introduction}

With the running coupling constant now a familiar concept already amply 
verified in experiment, one would not be surprised that the fermion mass 
matrix also varies with changing energy scales.  That its eigenvalues, 
namely the fermion masses, actually do run has already been experimentally 
verified in certain circumstances \cite{Santamaria}.  What we consider in 
this paper is the scenario when the fermion mass matrix changes also its 
orientation (rotates) in generation space as the scale changes.

Theoretically, of course, there are good reasons to expect that the fermion
mass matrix will rotate with changing scales.  Even in the Standard Model 
mass matrix rotation occurs so long as there is nontrivial mixing 
between the up and down fermion states.  The renormalization group equation 
there satisfied by the mass matrix $U$ of the up fermions \cite{Grzadsen}:
\begin{equation}
16 \pi^2 \frac{d U}{d t} = -\frac{3}{2} D D^{\dagger} U + ...
\label{rgeqsm}
\end{equation}
contains already at leading order
a term on the right which is nondiagonal in the eigenstates of
$U$ when the mass matrix $D$ of the down fermions is related to $U$ by a
nontrivial mixing matrix, so that a $U$ matrix diagonalized at one scale
can no longer remain diagonal at another scale, or in other words, it will
rotate as claimed.  The same argument holds for the mass matrix $D$ for 
the down fermions.  This fact was pointed out for quarks already long 
ago in, for example, \cite{Ramond}, and now that nontrivial mixing for 
leptons has been strongly indicated \cite{Superk,Soudan,Homestake,Gallex,
Sage,Sno}, if not already confirmed by experiment, the same conclusion 
can be drawn for leptons also.  Next, looking beyond the present Standard 
Model, one encounters further possible mechanisms for driving the mass 
matrix rotation.  Indeed, the mere fact that different generations of 
fermions can rotate into one another, as the mass matrix rotation implies, 
already means that they are not distinct entities as once conceived but 
just different manifestations of the same object, like the different 
colours of a quark, related presumably by some ``horizontal'' symmetry 
\cite{horsym}.  This suggests new forces which can change the generation 
index, and hence drive directly mass matrix rotations, in addition and 
in contrast to the ``indirect'' mechanism in (\ref{rgeqsm}) via mixing.  

However, even apart from theoretical prejudices as above indicated, there 
are, in our opinion, already some empirical indications which argue strongly, 
though perhaps as yet only circumstantially, in favour of fermion mass 
matrix rotation.  These come about as follows.  As is well-known, quarks 
and leptons exhibit remarkable mass and mixing patterns.  First, both 
their mass spectra are hierarchical, with the masses falling by one to two 
orders in magnitude from generation to generation.  Secondly, the mixing 
matrices (i.e. CKM \cite{Cabibbo,KM} for quarks and MNS \cite{MNS} for 
leptons) which parametrize the relative orientations between up and down 
states seem roughly similar in shape for quarks \cite{databook} and leptons, 
at least as far as is already known about the latter from the data 
in neutrino 
oscillation \cite{Superk,Soudan,Homestake,Gallex,Sage,Sno}, only with 
the off-diagonal elements generally much larger for leptons than for
quarks .  Neither of these features have any explanation in the present 
Standard Model, in which they are just taken for granted, and between them 
they account for some two-thirds of the Standard Model's 20-odd empirical 
(``fundamental'') parameters.  However, as was pointed out in a recent
note \cite{empirdsm}, if one assumes that the fermion mass matrix rotates 
at a certain speed, then these features can all be very simply understood.  

That this is the case can be summarized as follows.  Once the mass matrix 
is allowed to rotate with changing scales, then the usual definition of 
fermion flavour states as its eigenstates will have to be refined since, 
the eigenstates being now scale-dependent, one has to specify at what 
scale(s) the fermion flavour states are taken as the eigenstates.  If 
the fermion masses are defined as the eigenvalues of the mass matrix at 
the scales equal in value to the masses themselves, as is usually done, then 
it would be natural to define the corresponding fermion flavour states as the 
eigenstates at the same scales also.  It follows then that the state vectors 
of, say, the $t$ and $ b$ quarks will be defined at different scales, 
so that even if the $U$- and $D$-quark mass matrices share always the 
same orientation at the same scale, the rotation of the mass matrix from 
the scale of the $t$ mass to that of the $b$ mass will already imply a 
disorientation between the $t$ and $b$ state vectors, or in other words 
a CKM matrix element $V_{tb}$ represented by the direction cosine between 
the $t$ and $b$ vectors which is different from unity.  More generally, 
it can be seen along the same lines that a rotating mass matrix will 
generate not only a nontrivial mixing (CKM or MNS) matrix but also 
nonzero masses for lower generation fermions even when one starts with 
neither.  Indeed, what was shown in \cite{empirdsm} was that all existing 
empirical information on the fermion mass hierarchy and mixing pattern, 
(excepting for the moment only $CP$-violation) can now be understood 
as consequences a rotating mass matrix in the above manner, at least 
qualitatively but in some cases even quantitatively.  We regard this as 
a rather strong though indirect empirical indication in favour of 
fermion mass matrix rotation.  Indeed, if this interpretation of fermion 
mixing and mass hierarchy is accepted, then it implies a rotation speed
considerably greater than that driven by the mechanism (\ref{rgeqsm}) 
in the current Standard Model framework and suggests a driving mechanism
from beyond that.

Given these indications, both empirical and theoretical, it would be 
natural to enquire what other physical implications a rotating fermion 
mass matrix may have which can be tested directly by experiment.  One obvious
candidate is flavour-violation, for a rotating mass matrix will not remain
diagonal in the flavour states at scales other than the scale(s) at which
these flavour states are defined.  And since reaction amplitudes depend in
general on the fermion mass matrices, these too can become flavour-nondiagonal
leading thus to flavour-violating reactions.  This possibility has already 
been considered in general terms in, for example, \cite{impromat} in some 
detail.  We need here only to give an outline of 
its physical significance.  

The importance of flavour-conservation as a possible fundamental concept,
of course, has long been recognized and its consequences subjected to 
rigorous experimental tests.  In particular, the very stringent bounds set 
on $\mu \longrightarrow e \gamma$ and $\mu \longrightarrow e e \bar{e}$
decays, which are at present respectively $1.2 \times 10^{-11}$ and
$1.0 \times 10^{-12}$ for the branching ratio over the predominant, but already
weak, decay for $\mu$, tend to give the impression that any violation of
lepton flavour would have to be extremely small.  However, such a conclusion
may be premature for not having taken account of the possibility that the 
lepton mass matrix rotates with the energy scale.  If the mass matrix does
rotate, then the fact that flavour-conservation has been stringently tested
in $\mu \longrightarrow e \gamma$ and $\mu \longrightarrow e e \bar{e}$
at the $\mu$ mass scale, where the $\mu$ state is by definition diagonal,
is by itself no guarantee that flavour-violation will
be equally small in another reaction at another scale where the mass 
matrix may have rotated to another orientation so that the $\mu$ state
is no longer diagonal there.  Such flavour-violating effects can in 
principle occur by virtue of the mass matrix rotation even when there 
are no explicit flavour-changing neutral current (FCNC) couplings in the  
action, and can be sizeable even when FCNC effects are small.  We have 
therefore suggested for them the term ``transmutation'' in \cite{impromat}
for distinction, which we shall adopt also in this paper.  The size of 
transmutation effects, if any are observed, and their variation with 
energy will give indications on how the rotation of the mass matrix
is driven, the knowledge of which may in turn shed light on the origin 
of generations itself, a basic question in particle physics that has 
already been with us for many years.  For this reason, we suggest that
flavour-violation be routinely tested in experiment whenever conditions
are favourable. 

For pursuing this program, a method was developed in \cite{photrans} for 
calculating the flavour-violation effects due just to the kinematics of 
a rotating mass matrix.  In this, one treats the rotating mass matrix as
a proposition in isolation without enquiring from what mechanism this 
rotation originates and without taking account of other possible rotation 
effects which might in principle accompany the mass matrix rotation.  
Such a preliminary attitude, we think, is reasonable given that the only 
empirical evidence one has so far is for the rotating mass matrix alone 
\cite{empirdsm} with no direct hints yet of the mechanism driving it.  The 
result of such a calculation will serve to gauge what size flavour-violation 
effects might in principle be expected from a mass matrix rotating at a 
given speed with the aim of providing an indicator for experimenters 
planning an analysis along these line.  However, it is not to be taken 
as a necessary prediction of a rotating mass matrix under general 
circumstances for the following reason.  Reaction amplitudes depend on 
quantities other than the fermion mass matrix, such as, say, interaction 
vertices, which may in principle also rotate, and any theoretical mechanism 
for driving the mass matrix rotation may also imply vertex rotations.  And
these other rotational effects may modify or even cancel the effects computed 
by the above method from the kinematics of the rotating mass matrix alone.  
Indeed, in a detailed calculation performed specifically in the Dualized 
Standard Model (DSM) framework that we ourselves suggeted which is  
reported in a separate paper \cite{transmudsm}, such a cancellation is 
found in fact to occur.  Hence, calculations done with the method of 
\cite{impromat,photrans} as those in the the present paper have to be taken 
with this reservation in mind, which reservation was unfortunately not 
made clear in the earlier references because it was not clear then even 
to ourselves. 

In the present paper, we choose to investigate transmutation effects 
in the following reactions by the method suggested in \cite{photrans}
which, though originally developed for photo-transmutation, can be adapted
to the present case with but minor modifications:
\begin{equation}
e^+ e^- \longrightarrow e^+ \tau^-, \tau^+ e^-;
\label{etotau}
\end{equation}
\begin{equation}
e^+ e^- \longrightarrow e^+ \mu^-, \mu^+ e^-;
\label{etomu}
\end{equation}
\begin{equation}
e^+ e^- \longrightarrow \mu^+ \tau^-, \tau^+ \mu^-.
\label{eetomutau}
\end{equation}
The obvious practical reason for investigating these reactions is that 
there are several high intensity machines in operation, such as BEPC, CESR 
(Cleo), PEP II (BaBar), and KEK II (Belle), which appear capable of observing 
these flavour-violating effects to high accuracy, besides LEP, which though 
now turned off, has left still masses of data which could be useful 
for the same purpose.  

\begin{figure}[ht]
\begin{center}
{\unitlength=1.0 pt \SetScale{1.0} \SetWidth{1.0}
\begin{picture}(350,130)(0,0)

\Photon(65,25)(65,75){3}{6.5}
\ArrowLine(65,75)(30,100)
\ArrowLine(30,0)(65,25)
\ArrowLine(65,25)(100,0)
\ArrowLine(100,100)(65,75)

\Text(20,0)[]{$\ell_\alpha^-$}
\Text(20,100)[]{$e^+$}
\Text(110,100)[]{$e^+$}
\Text(110,0)[]{$\ell_\beta^-$}
\Text(40,20)[]{$p$}
\Text(40,80)[]{$q$}
\Text(90,22)[]{$p'$}
\Text(90,80)[]{$q'$}
\Text(75,50)[]{$\gamma$}
\Text(65,-20)[]{$(a)$}

\Photon(230,50)(300,50){3}{6.5}
\ArrowLine(230,50)(200,80)
\ArrowLine(330,80)(300,50)
\ArrowLine(200,20)(230,50)
\ArrowLine(300,50)(330,20)

\Text(190,20)[]{$e^-$}
\Text(340,20)[]{$\ell_\beta^-$}
\Text(190,80)[]{$e^+$}
\Text(340,80)[]{$\ell_\alpha^+$}
\Text(225,30)[]{$p$}
\Text(305,30)[]{$p'$}
\Text(225,70)[]{$q$}
\Text(305,70)[]{$q'$}
\Text(265,65)[]{$\gamma$}
\Text(265,-20)[]{$(b)$}
\end{picture}}
\end{center}
\vspace*{5mm}
\caption{Feynman diagrams for the transmutation amplitude.}
\label{Feyndi}
\end{figure}
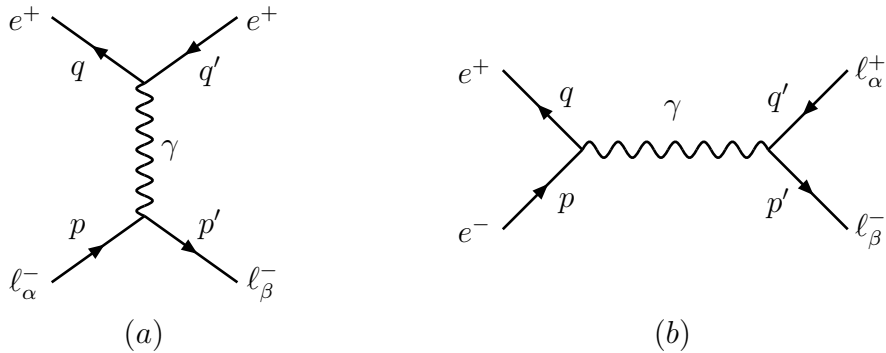

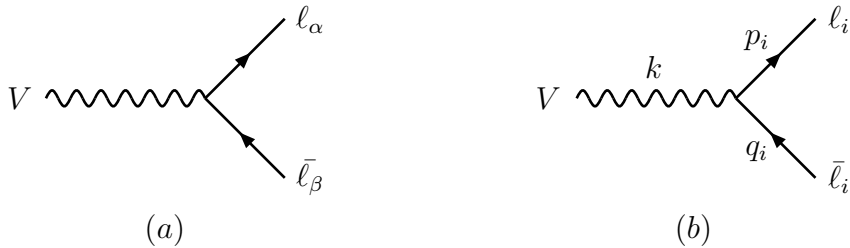
\begin{figure}[ht]
\begin{center}
{\unitlength=1.0 pt \SetScale{1.0} \SetWidth{1.0}
\begin{picture}(350,100)(-20,0)
\Photon(0,30)(60,30){3}{6.5}
\ArrowLine(60,30)(90,60)
\ArrowLine(90,0)(60,30)

\Text(-10,30)[]{$V$}
\Text(100,60)[]{$\ell_\alpha$}
\Text(100,0)[]{$\bar{\ell_\beta}$}
\Text(45,-20)[]{$(a)$}

\Photon(200,30)(260,30){3}{6.5}
\ArrowLine(260,30)(290,60)
\ArrowLine(290,0)(260,30)
\Text(190,30)[]{$V$}
\Text(230,42)[]{$k$}
\Text(269,52)[]{$p_i$}
\Text(269,10)[]{$q_i$}
\Text(300,60)[]{$\ell_i$}
\Text(300,0)[]{$\bar{\ell_i}$}
\Text(245,-20)[]{$(b)$}
\end{picture} }
\end{center}
\vskip 0.5cm
\caption{Decay amplitudes}
\label{decaydiag}
\end{figure}

Lepton transmutation in reactions 
(\ref{etotau})---(\ref{eetomutau}) can proceed by, for example, the processes 
represented by the Feynman diagrams in Figure \ref{Feyndi}.  At the energy 
at which an experiment is performed, the amplitudes for these processes, 
being dependent on the lepton masses, are diagonal in the eigenstates 
$j = 1,2,3$ of the mass matrix at that scale but not in general, by the 
reasoning above, diagonal in the flavour states $e, \mu, \tau$.  And this 
fact alone could be enough to give lepton transmutation as a result.  The 
transmutation reactions (\ref{etotau})---(\ref{eetomutau}) can of course 
occur also via other processes, such as higher order photon-exchange diagrams 
or $Z_0$-exchange, but the effects from these for the energy range of 
present interest are small and will be neglected.  What can give sizeable 
contributions, however, is the formation of vector bosons in the intermediate 
state followed by their subsequent (transmutational) decays, as represented 
by Figure \ref{decaydiag} (b).

Figures \ref{Feyndi} and \ref{decaydiag}, then, are the only processes 
we shall consider in this paper, where we shall show that, by following 
the procedure suggested in \cite{photrans}, one can calculate explicitly 
the differential cross sections for the 3 transmutational reactions 
(\ref{etotau}), (\ref{etomu}), and (\ref{eetomutau}), given any rotating 
mass matrix.  The rotating mass matrix itself will figure only in the 
normalization of the cross sections, not in their angular or spin dependence 
both of which are given essentially just by kinematics. 

As a numerical example for the sort of cross sections one might expect
for these transmutation reactions, let us consider in particular the 
reaction (\ref{eetomutau}).  The normalization of the cross section of
this reaction is given by the rotation angle between the $\mu$ and $\tau$ 
states at the energy scale of the experiment, a good estimate for which
can be read already from the existing data on fermion mass and mixing 
patterns interpreted as rotation effects as in \cite{empirdsm}, say from
Figure 3 of that paper.  Specifically, for $\sqrt{s} = 10.58$ GeV at the 
mass of the $\Upsilon(4S)$ at which BaBar is run, one obtains in this way
an estimate for the integrated cross section of reaction (\ref{eetomutau})
of around 80 fb which translates to as many as a thousand events in the 
data they have already collected from their first year's run of 20 
${\rm fb}^{-1}$, assuming 100 percent detection efficiency, and if so 
should be readily detectable.  

However, as already stressed above, these estimates for transmutation 
cross sections are not to be interpreted as definitive predictions for 
the quoted rotating mass matrix since they can be modified by other rotation
effects.  Nevertheless, they are of interest in giving an idea of the 
size of flavour-violation effects that a rotating mass matrix can in 
principle generate.  Given the smallness of flavour-violation at the $\mu$ 
mass scale in $\mu \rightarrow e \gamma$ and $\mu \rightarrow e e e$ decay, 
any detection of flavour-violation in the reactions 
(\ref{etotau})---(\ref{eetomutau}) at BaBar or similar experiment 
at a different scale would 
be a positive indication for mass matrix rotation, although a negative result, 
by virtue of the preceding observation, would not be able at present to rule 
it out.

\section{The Reaction Amplitude (a)}

\begin{figure}[ht]
\begin{center}
{\unitlength=1.0 pt \SetScale{1.0} \SetWidth{1.0}
\begin{picture}(350,130)(0,0)

\Photon(65,25)(65,75){3}{6.5}
\ArrowLine(65,75)(30,100)
\ArrowLine(30,0)(65,25)
\ArrowLine(65,25)(100,0)
\ArrowLine(100,100)(65,75)

\Text(20,0)[]{$\ell_j^-$}
\Text(20,100)[]{$e^+$}
\Text(110,100)[]{$e^+$}
\Text(110,0)[]{$\ell_j^-$}
\Text(40,20)[]{$p_j$}
\Text(40,80)[]{$q$}
\Text(90,22)[]{${p'}_j$}
\Text(90,80)[]{$q'$}
\Text(75,50)[]{$\gamma$}
\Text(65,-20)[]{$(a)$}

\Photon(230,50)(300,50){3}{6.5}
\ArrowLine(230,50)(200,80)
\ArrowLine(330,80)(300,50)
\ArrowLine(200,20)(230,50)
\ArrowLine(300,50)(330,20)

\Text(190,20)[]{$e^-$}
\Text(340,20)[]{$\ell_j^-$}
\Text(190,80)[]{$e^+$}
\Text(340,80)[]{$\ell_j^+$}
\Text(225,30)[]{$p$}
\Text(305,30)[]{${p'}_j$}
\Text(225,70)[]{$q$}
\Text(305,70)[]{${q'}_j$}
\Text(265,65)[]{$\gamma$}
\Text(265,-20)[]{$(b)$}
\end{picture}}
\end{center}
\vspace*{5mm}
\caption{Feynman diagram for the diagonal amplitudes.}
\label{Feyndiag}
\end{figure}
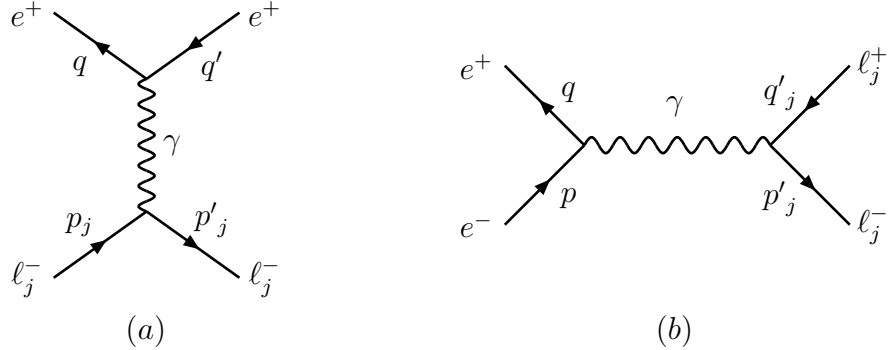

Consider first the one-photon exchange diagram of Figure \ref{Feyndi}(a),
which will be seen to give the dominant contribution to the two reactions 
(\ref{etotau}) and (\ref{etomu}).  It does not contribute to the reaction
(\ref{eetomutau}), which can proceed by one-photon exchange in $e^+e^-$ 
collision only when both $e^+$ and $e^-$ transmute, but this will be so 
far down in magnitude as to be negligible for present consideration.  
Then according to the procedure suggested in \cite{photrans}, at any 
given energy $\sqrt{s}$, the transmutation amplitude for the reaction :
\begin{equation}
e^+ \ell^-_\alpha \longrightarrow e^+ \ell^-_\beta,
\label{atob}
\end{equation}
is given by a rotation in generation space, thus:
\begin{equation}
{\cal M}^{(a)} = \sum_j S^\dagger_{\beta j}{\cal M}^{(a)}_j 
   S_{\alpha j},
\label{calMa}
\end{equation}
from the diagonal amplitudes ${\cal M}^{(a)}_j$ for the reaction:
\begin{equation}
e^+ \ell^-_j \longrightarrow e^+ \ell^-_j
\label{jtoj}
\end{equation}
for the mass eigenstate $j$ at the scale $\sqrt{s}$ with eigenvalue $m_j$, 
where $S_{\alpha j} = \langle j | \alpha \rangle$ is the rotation matrix 
in generation space which relates the triad of lepton flavour states 
$\alpha = e, \mu, \tau$ to the eigentriad $j$ at the scale $\sqrt{s}$.  
Explicitly, for the one-photon exchange diagram of Figure \ref{Feyndiag}(a), 
we have:
\begin{equation}
({\cal M}^{(a)}_j)^{r'r}_{s's} = - i e^2 [\bar{u}_{s'}(p'_j) \gamma^\mu
   u_s(p_j)] \frac{1}{(p_j' - p_j)^2} [\bar{v}_r(q) \gamma_\mu v_{r'}(q')],
\label{calMaj}
\end{equation}
where $s$ and $s'$ denote the spins of the incoming and outgoing
lepton $\ell^-$ and $r$ and $r'$ those of the antileptons $\ell^+$.

To actually evaluate (\ref{calMaj}), we have still to specify the values of 
the momenta $p_j$ and $p_j'$ entering there.  The point is that the amplitude 
for the two-body reaction (\ref{jtoj}) is of course a function of only two 
variables, which we may take to be the standard Mandelstam variables $s$ 
and $t$, so that all components of the various momenta appearing in 
(\ref{calMaj}) must be expressible in terms of them.  The reasoning 
required to arrive at these expressions is not entirely trivial, but 
following the considerations given in \cite{photrans} which apply as well 
to the present case with but minor modifications, we obtain the following 
relationships between the different momenta to be used later for 
deriving the required expressions:
\begin{eqnarray}
p_j & = & a_j q +  b_j q' + c_j p_i \nonumber \\
p_j' & = & a_j q' +  b_j q + c_j p_i',
\label{reqsandps}
\end{eqnarray}
where
\begin{eqnarray}
c_j & = & \sqrt{\frac{(s - m_j^2 + m^2)^2 + st_0}{(s - m^2)^2 + st}}, 
   \nonumber \\
b_j & = & \frac{c_j(s + m^2 ) - (s - m_j^2 + m^2 )}{t_0},
   \nonumber \\
a_j & = & \frac{c_j(s + m^2+ t_0) - (s - m_j^2 + m^2 + t_0)}{t_0},
\label{abc}
\end{eqnarray}
with $t_0 = t - 4m^2$, $m$ the positron mass, and $m_i$ put equal to zero
for reasons to be made apparent.
 
In this paper, we shall be interested mainly in unpolarized cross sections 
which means that we shall need to evaluate sums of the absolute values 
squared of the amplitudes (\ref{calMa}) over all the spins $s,r,s',r'$.  
These spin-sums, as was explained in \cite{photrans}, are not so readily 
performed as usual by the standard method of taking traces of 
$\gamma$-matrices because of the crossed terms between channels labelled 
by different $j$'s obtained in squaring (\ref{calMa}).  We shall therefore 
follow the tactics adopted in \cite{photrans} of explicitly performing the 
spin-sums in a specific Lorentz frame with a specific representation of 
the $\gamma$-matrices.  As in \cite{photrans}, we choose to work in the cm 
frame of the channel $i = 3$ which in our convention denotes the mass 
eigenstate with the lowest mass $m_3$, which, being in all cases considered
at most of the order of the electron mass and therefore negligible, is put 
equal to zero.  This gives:
\begin{eqnarray}
q^\mu & = & (E, 0, 0, \omega), \nonumber \\
q'^\mu & = & (E, 0, \omega \sin \theta_3', \omega \cos \theta_3'), 
   \nonumber \\    
p_3^\mu & = & (\omega, 0, 0, -\omega), \nonumber \\
p_3'^\mu & = & (\omega, 0, -\omega \sin \theta_3', -\omega 
   \cos \theta_3'), \nonumber \\
p_j^\mu & = & (E_j, 0, - \omega_j \sin \theta_j, - \omega_j 
   \cos \theta_j), \nonumber \\
p_j'^\mu & = & (E_j, 0, - \omega_j \sin \theta_j', - \omega_j
   \cos \theta_j'),
\label{qsandps}
\end{eqnarray}
with 
\begin{eqnarray}
E & = & \frac{s + m^2}{2 \sqrt{s}}, \nonumber \\
\omega & = & \frac{s - m^2}{2 \sqrt{s}}, \nonumber \\
\cos \theta_3' & = & 1 + \frac{2 s t}{(s - m^2)^2}.
\label{Eomcos}
\end{eqnarray}
Further, from (\ref{reqsandps})--(\ref{qsandps}), one obtains
\begin{eqnarray}
E_j &=& \frac{1}{2 \sqrt{s} t_0} \Big\{2\sqrt{[(s - m^2)^2 + st]
   [(s - m_j^2 + m^2)^2 + st_0]}  \nonumber \\ 
&&  \mbox{} -  (s + m^2)[2(s - m^2 - m_j^2) 
   + t]\Big\},
\label{Ej}
\end{eqnarray}
with
\begin{equation}
\omega_j = \sqrt{E_j^2 - m_j^2},
\label{omegaj}
\end{equation}
and
\begin{eqnarray}
\sin \theta_j & = & \frac{-\sqrt{-st}}{2 s t_0 \omega \omega_j}
   \Big\{(s + m^2) \sqrt{(s - m_j^2 + m^2)^2 + s t_0} \nonumber \\
&& \mbox{} - (s - m_j^2 +m^2)
   \sqrt{(s - m^2)^2 + st}\Big\}, \nonumber \\
\sin \theta_j' & = & \frac{-\sqrt{-st}}{2 s t_0 \omega \omega_j}
   \Big\{(s + m^2) \sqrt{(s - m_j^2 + m^2)^2 + s t_0} \nonumber \\
&& \mbox{} - (s - m_j^2 +m^2
   + t_0) \sqrt{(s - m^2)^2 + st}\Big\},
\label{sinsinp}
\end{eqnarray}
with
\begin{equation}
\theta_3' = \theta_j + \theta_j'.
\label{sumtheta}
\end{equation}
These formulae (\ref{Eomcos})--(\ref{sinsinp}), together with the formulae
in the last paragraph, all reduce to the corresponding formulae derived in 
\cite{photrans} for photo-trans\-muta\-tion if we put the mass $m$ of the 
$e^+$ in (\ref{jtoj}) equal to zero, which will indeed be a very good 
approximation in most applications.  We have kept the dependence on $m$ 
explicit only for the sake of generality in case the formulae are to be 
applied in future to other circumstances, such as lepton transmutations 
in $\mu^+ \mu^-$ collisions.

We choose again for $\gamma$-matrices the Pauli--Dirac representation:
\begin{equation}
\gamma^0 = \left( \begin{array}{cc} 1 & 0 \\ 0 & -1 \end{array} \right); \ 
\gamma^k = \left( \begin{array}{cc} 0 & \sigma_k \\ -\sigma_k & 0 
   \end{array} \right).
\label{gammarep}
\end{equation}
The spins of the incoming and outgoing leptons $j$ we quantize along the
direction $p_3$ and $p_3'$ respectively, while the spins of the $e^+$,
whether incoming or outgoing, we quantize along its direction of motion.
With these specifications, the wave functions of the leptons $j$ are
given by:
\begin{eqnarray}
u_+(p_j) & = & \frac{1}{\sqrt{2(E_j+m_j)}} \left( \begin{array}{c}
   0 \\ E_j + m_j \\ 0 \\ \omega_j e^{-i \theta_j} \end{array} \right),
   \nonumber \\  
u_-(p_j) & = & \frac{1}{\sqrt{2(E_j+m_j)}} \left( \begin{array}{c}
   E_j + m_j \\ 0 \\ -\omega_j e^{-i \theta_j} \\ 0 \end{array} \right),
\label{upj}
\end{eqnarray}
and
\begin{eqnarray}
u_+(p_j') & = & \frac{1}{2 \sqrt{2(E_j+m_j)}} \left( \begin{array}{c}
   (E_j + m_j)(1 - e^{-i \theta_3'}) \\ (E_j + m_j)(1 + e^{-i \theta_3'}) 
   \\ \omega_j (e^{i \theta_j} - e^{-i \theta_j'}) \\ \omega_j 
   (e^{i \theta_j} + e^{-i \theta_j'}) \end{array} \right), \nonumber \\ 
u_-(p_j') & = & \frac{1}{2 \sqrt{2(E_j+m_j)}} \left( \begin{array}{c}
   (E_j + m_j)(1 + e^{-i \theta_3'}) \\ (E_j + m_j)(1 - e^{-i \theta_3'}) 
   \\ - \omega_j (e^{i \theta_j} + e^{-i \theta_j'}) \\ - \omega_j 
   (e^{i \theta_j} - e^{-i \theta_j'}) \end{array} \right); 
\label{upjp}
\end{eqnarray}
while for the $e^+$, we have the wave functions:
\begin{eqnarray}
v_+(q) & = & \frac{1}{\sqrt{2(E + m)}} \left( \begin{array}{c}
   0 \\ - \omega \\ 0 \\ E + m \end{array} \right),
   \nonumber \\  
v_-(q) & = & \frac{1}{\sqrt{2(E + m)}} \left( \begin{array}{c}
   \omega \\ 0 \\ E + m \\ 0 \end{array} \right),
\label{vq}
\end{eqnarray}
and:
\begin{eqnarray}
v_+(q') & = & \frac{1}{2 \sqrt{2(E + m)}} \left( \begin{array}{c}
   - \omega (1 - e^{-i \theta_3'}) \\ - \omega (1 + e^{-i \theta_3'}) \\
   (E + m)(1 - e^{-i \theta_3'}) \\ (E + m)(1 + e^{-i \theta_3'}) 
   \end{array} \right), \nonumber \\
v_-(q') & = & \frac{1}{2 \sqrt{2(E + m)}} \left( \begin{array}{c}
   \omega (1 + e^{-i \theta_3'}) \\ \omega (1 - e^{-i \theta_3'}) \\
   (E + m)(1 + e^{-i \theta_3'}) \\ (E + m)(1 - e^{-i \theta_3'})
   \end{array} \right).
\label{vqp}
\end{eqnarray}

With the wave functions in (\ref{upj})--(\ref{vqp}), it is straightforward 
to evaluate the diagonal amplitudes in (\ref{calMaj}).  We obtain the
following:
\begin{eqnarray}
({\cal M}^{(a)}_j)^{++}_{++} & = & ({\cal M}^{(a)}_j)^{--}_{--} \nonumber \\
   & = & - \frac{i e^2}{t} \Big\{\frac{E}{4} (1 + \cos \theta_3')
         [(E_j + m_j) + (E_j - m_j) e^{-2 i \theta_j}] \nonumber \\
   &   & + \omega \omega_j (1 - \cos \theta_3') e^{-i \theta_j}
         + \frac{\omega \omega_j}{2}(1 + \cos \theta_3') e^{-i \theta_j}
         \Big\} \nonumber \\
({\cal M}^{(a)}_j)^{--}_{++} & = & ({\cal M}^{(a)}_j)^{++}_{--} \nonumber \\
   & = & - \frac{i e^2}{t} (1 + \cos \theta_3') \Big\{ \frac{E}{4}
         [(E_j + m_j) + (E_j - m_j) e^{-2 i \theta_j}] \nonumber \\
   &   & + \frac{\omega \omega_j}{2} e^{-i \theta_j} \Big\} \nonumber \\
({\cal M}^{(a)}_j)^{+-}_{+-} & = & ({\cal M}^{(a)}_j)^{-+}_{-+} =
   ({\cal M}^{(a)}_j)^{-+}_{+-} = ({\cal M}^{(a)}_j)^{+-}_{-+} \nonumber \\
   & = & - \frac{i e^2}{t} \frac{m}{4}(1 - \cos \theta_3')
       \Big\{(E_j + m_j) - (E_j - m_j) e^{-2 i \theta_j}\Big\} \nonumber \\
({\cal M}^{(a)}_j)^{+-}_{++} & = & ({\cal M}^{(a)}_j)^{-+}_{++} =
   ({\cal M}^{(a)}_j)^{+-}_{--} = ({\cal M}^{(a)}_j)^{-+}_{--} \nonumber \\
   & = & \frac{e^2}{t} \frac{m}{4} \sin \theta_3'
       \Big\{(E_j + m_j) + (E_j - m_j) e^{-2 i \theta_j}\Big\} \nonumber \\
({\cal M}^{(a)}_j)^{++}_{+-} & = & ({\cal M}^{(a)}_j)^{++}_{-+} =
   ({\cal M}^{(a)}_j)^{--}_{+-} = ({\cal M}^{(a)}_j)^{--}_{-+} \nonumber \\
   & = & - \frac{e^2}{t} \frac{E}{4} \sin \theta_3'
       \Big\{(E_j + m_j) - (E_j - m_j) e^{-2 i \theta_j}\Big\},
\label{calMajsp}
\end{eqnarray}
where subscripts denote the spins of the leptons $j$ and superscripts the
spins of the $e^+$, with the right index pertaining to the incoming and
the left index to the outgoing particle.  Using the formulae derived 
earlier in (\ref{Eomcos})--(\ref{sinsinp}), these amplitudes can then
be expressed in terms of the Mandelstam invariants $s$ and $t$ as desired.

\section{The Reaction Amplitude (b)}

Turning next to the diagram of Figure \ref{Feyndi}(b) which contributes to
all three reactions (\ref{etotau})--(\ref{eetomutau}), we proceed in a
similar manner.  The transmutational amplitude for the reaction:
\begin{equation}
e^+ e^- \longrightarrow \ell^+_\alpha \ell^-_\beta,
\label{eetoab}
\end{equation}
is given by a rotation in generation space, thus:
\begin{equation}
{\cal M}^{(b)} = \sum_j S_{\alpha j} S^\dagger_{\beta j}
   {\cal M}^{(b)}_j,
\label{calMb}
\end{equation}
from the diagonal amplitudes ${\cal M}^{(b)}_j$ depicted in Figure 
\ref{Feyndiag}(b) for the reaction:
\begin{equation}
e^+ e^- \longrightarrow \ell^+_j \ell^-_j
\label{eetojj}
\end{equation}
where
\begin{equation}
({\cal M}^{(b)}_j)^{r'r}_{s's} = i e^2 [\bar{u}_{s'}(p'_j) \gamma^\mu
   v_{r'}(q'_j)] \frac{1}{(p + q)^2} [\bar{v}_r(q) \gamma_\mu u_s(p)].
\label{calMbj}
\end{equation}
We work now in the cm of the incoming $e^+$ and $e^-$ system with
\begin{eqnarray}
q^\mu & = & (E, 0, 0, \omega), \nonumber \\
p^\mu & = & (E, 0, 0, - \omega), \nonumber \\
{p'}^\mu_j & = & (E_j, 0, - \omega_j \sin {\theta'}_j, - \omega_j 
   \cos {\theta'}_j), \nonumber \\
{q'}^\mu_j & = & (E_j, 0, \omega_j \sin {\theta'}_j, \omega_j 
   \cos \theta_j'),
\label{pqb}
\end{eqnarray}
and
\begin{eqnarray}
E & = & E_j = \sqrt{s}/2, \nonumber \\
\omega & = & \sqrt{E^2 - m^2}; \ \ \omega_j = \sqrt{E^2 - m_j^2},
   \nonumber \\
\cos \theta_3' & = & 1 + \frac{t}{2 E^2}, \nonumber \\
\cos \theta_j' & = & \frac{t - m^2 - m_j^2 + 2 E^2}{2 \omega \omega_j}.
\label{Eomcosb}
\end{eqnarray}
Further, with the spins of the outgoing particles $\ell_j^+, \ell_j^-$
quantized along $q_3'$ and $p_3'$ respectively, and those of the
incoming $e^+, e^-$ along their directions of motion $q$ and $p$
respectively, the wave functions are given by:
\begin{eqnarray}
u_+(p) & = & \frac{1}{\sqrt{2(E+m)}} \left( \begin{array}{c}
   0 \\ E+m \\ 0 \\ \omega \end{array} \right), \nonumber \\
u_-(p) & = & \frac{1}{\sqrt{2(E+m)}} \left( \begin{array}{c}
   E+m \\ 0 \\ - \omega \\ 0 \end{array} \right), \nonumber \\
v_+(q) & = & \frac{1}{\sqrt{2(E+m)}} \left( \begin{array}{c}
   0 \\ - \omega \\ 0 \\ E+m \end{array} \right), \nonumber \\
v_-(q) & = & \frac{1}{\sqrt{2(E+m)}} \left( \begin{array}{c}
   \omega \\ 0 \\ E+m \\ 0 \end{array} \right),
\label{uvpq}
\end{eqnarray}
and
\begin{eqnarray}
u_+(p_j') & = & \frac{1}{2 \sqrt{2(E_j+m_j)}} \left( \begin{array}{c}
   (E_j + m_j)(1 - e^{-i \theta_j'}) \\ (E_j + m_j)(1 + e^{-i \theta_j'}) 
   \\ \omega_j (1 - e^{-i \theta_j'}) \\ \omega_j (1 + e^{-i \theta_j'}) 
   \end{array} \right), \nonumber \\ 
u_-(p_j') & = & \frac{1}{2 \sqrt{2(E_j+m_j)}} \left( \begin{array}{c}
   (E_j + m_j)(1 + e^{-i \theta_j'}) \\ (E_j + m_j)(1 - e^{-i \theta_j'}) 
   \\ - \omega_j (1 + e^{-i \theta_j'}) \\ - \omega_j (1 - e^{-i \theta_j'}) 
   \end{array} \right), \nonumber \\
v_+(q_j') & = & \frac{1}{2 \sqrt{2(E_j + m_j)}} \left( \begin{array}{c}
   - \omega_j (1 - e^{-i \theta_j'}) \\ - \omega_j (1 + e^{-i \theta_j'}) \\
   (E_j + m_j)(1 - e^{-i \theta_j'}) \\ (E_j + m_j)(1 + e^{-i \theta_j'}) 
   \end{array} \right), \nonumber \\
v_-(q_j') & = & \frac{1}{2 \sqrt{2(E_j + m_j)}} \left( \begin{array}{c}
   \omega_j (1 + e^{-i \theta_j'}) \\ \omega_j (1 - e^{-i \theta_j'}) \\
   (E_j + m_j)(1 + e^{-i \theta_j'}) \\ (E_j + m_j)(1 - e^{-i \theta_j'})
   \end{array} \right).
\label{uvpjqjp}
\end{eqnarray}
Hence, one obtains the amplitudes:
\begin{eqnarray}
({\cal M}^{(b)}_j)^{++}_{++} & = & ({\cal M}^{(b)}_j)^{--}_{--} 
   = - \frac{ie^2}{s}m m_j \cos \theta_j', \nonumber \\
({\cal M}^{(b)}_j)^{--}_{++} & = & ({\cal M}^{(b)}_j)^{++}_{--} 
   = -\frac{ie^2}{s}E E_j (1 + \cos \theta_j'), \nonumber \\
({\cal M}^{(b)}_j)^{+-}_{+-} & = & ({\cal M}^{(b)}_j)^{-+}_{-+} 
   = \frac{ie^2}{s}m m_j \cos \theta_j', \nonumber \\ 
({\cal M}^{(b)}_j)^{+-}_{++} & = & ({\cal M}^{(b)}_j)^{-+}_{--}
   =  \frac{e^2}{s}E m_j \sin \theta_j', \nonumber \\
({\cal M}^{(b)}_j)^{++}_{+-} & = & ({\cal M}^{(b)}_j)^{--}_{-+}
   = -\frac{e^2}{s} E m_j \sin \theta_j', \nonumber \\
({\cal M}^{(b)}_j)^{+-}_{--} & = & ({\cal M}^{(b)}_j)^{-+}_{++}
   = \frac{e^2}{s} m E_j \sin \theta_j', \nonumber \\
({\cal M}^{(b)}_j)^{-+}_{+-} & = & ({\cal M}^{(b)}_j)^{+-}_{-+} 
   = - \frac{ie^2}{s}E E_j (1 - \cos \theta_j') \nonumber \\
({\cal M}^{(b)}_j)^{++}_{-+} & = & ({\cal M}^{(b)}_j)^{--}_{+-} 
   = -\frac{e^2}{s} m E_j \sin \theta_j',
\label{calMbjsp}
\end{eqnarray}
which again can all be expressed in terms of the Mandelstam invariants 
$s$ and $t$ by means of the formulae in (\ref{Eomcosb}).

\section{Spin-summed Differential Cross Sections}

Substituting the diagonal amplitudes in (\ref{calMajsp}) and (\ref{calMbjsp})
into respectively the formulae (\ref{calMa}) and (\ref{calMb}) and adding 
the two contributions, one obtains the spin-amplitudes for the actual
transmutation reaction (\ref{etotau}), (\ref{etomu}) and (\ref{eetomutau}).
Hence, taking the absolute values squared of these amplitudes and summing
over all spins, one obtains the spin-summed differential cross sections as
desired:
\begin{equation}
\frac{d \sigma}{d \Omega} = \frac{1}{4\pi^2}\,\frac{1}{s}
   \,\frac{\omega'}{\omega} \,\frac{1}{4} \sum_{r,r';s,s'} 
   \big|({\cal M})^{r'r}_{s's} \big|^2 \times 0.3894\ {\rm mb/sr},
\label{crosssec}
\end{equation}
for energies measured in GeV, where $\omega$ and $\omega'$ are respectively 
the cm momenta of the actual incoming and outgoing particles in the 
transmutation reaction.  Although this is in principle straightforward, 
a few practical observations are in order.

First, in performing the sum in (\ref{calMa}) and (\ref{calMb}) over the 
diagonal states $j$, it is convenient to make use of the unitary property
of the rotation matrix $S_{\alpha j}$ to write, for $\alpha \neq \beta$:
\begin{equation}
\sum_j S_{\alpha j} S^\dagger_{\beta j} {\cal M}_j 
   = S_{\alpha 1} S^\dagger_{\beta 1} [{\cal M}_1 - {\cal M}_3],
\label{calMrewrite}
\end{equation}
which, as explained in \cite{photrans}, is a good approximation whenever 
the mass eigenvalues $m_j$ are hierarchical and avoids the need to know the
rotation matrix to unreasonably high accuracy.  Besides, as we shall see,
it gives us a clearer picture of how transmutation cross sections behave 
as functions of the Mandelstam invariants $s$ and $t$. 

\begin{figure}
\centerline{\psfig{figure=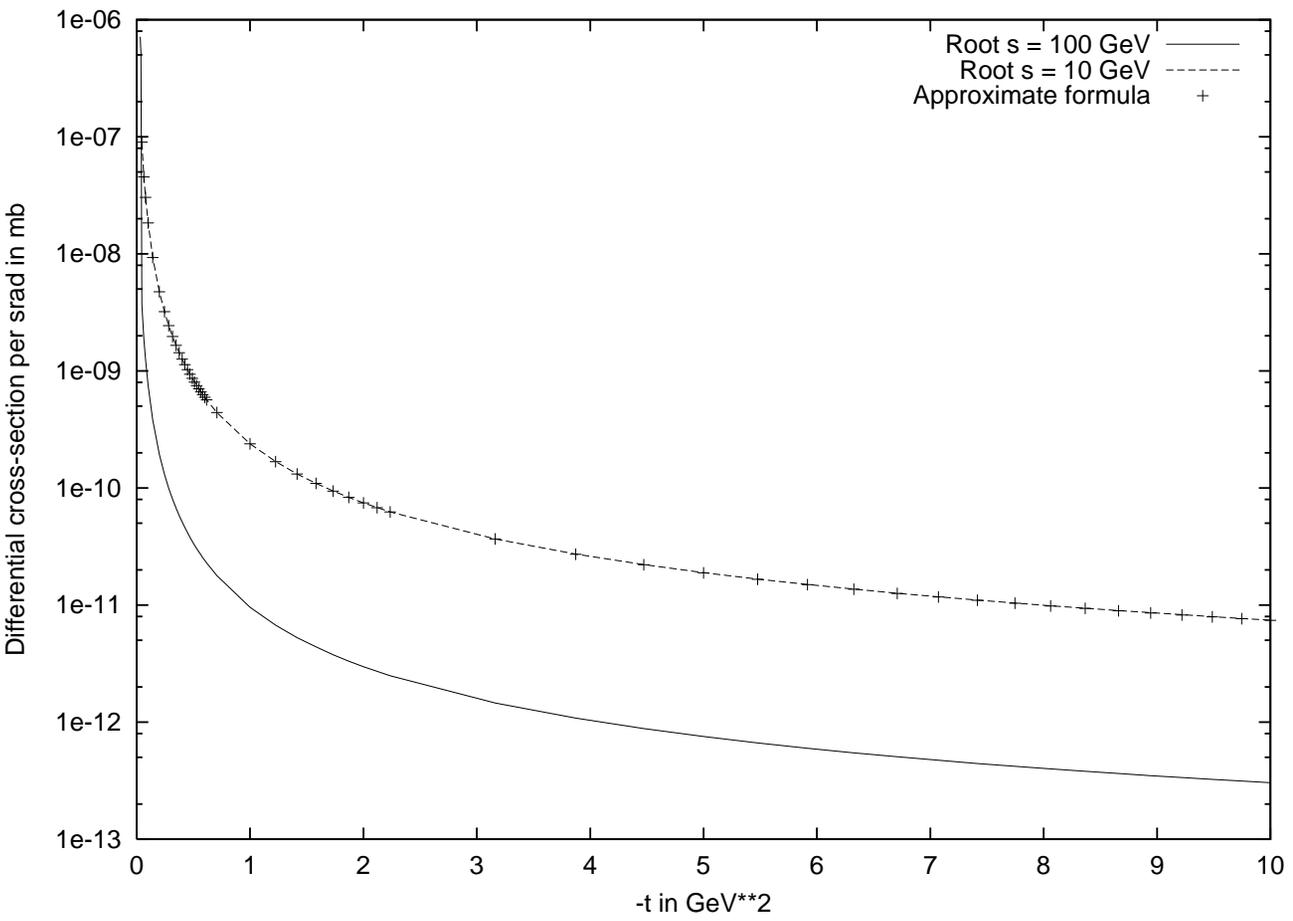,width=0.9\textwidth}}
\caption{Spin-summed differential cross sections for the reaction
$e^+ e^- \longrightarrow e^+ \tau^-$ at $\sqrt{s} =$ 10 GeV, 100 GeV.
For details see text.} 
\label{xsedexam}
\end{figure}

Second, as can be seen in (\ref{calMrewrite}), the transmutation amplitude
being proportional to ${\cal M}_1 - {\cal M}_3$ with the two amplitudes
differing just by the mass values, i.e. whether $m_1$ or $m_3$, the cross
section for transmutation is at most of order $m_1^2/s$ and decreases
rapidly with increasing energy.  For high $s$, therefore, ${\cal M}_1$
and ${\cal M}_3$ will largely cancel leading potentially to inaccuracy in
a direct calculation with the formula (\ref{calMrewrite}).  Indeed, this
was exactly what we found in our actual calculations, especially in the
the amplitude (a) where other large cancellations occur in the exact 
formulae.  This computational difficulty can be avoided just by expanding 
the amplitudes to order $m_1^2/s$ giving:
\begin{eqnarray}
({\cal M}^{(a)})^{++}_{++} & = & ({\cal M}^{(a)})^{--}_{--} \nonumber \\
   & = & \frac{i e^2}{t} \,\bigg\{\frac{m_1^2}{2 s} (s-t) - i m_1^2 
   \sqrt{\frac{-t}{s+t}}\bigg\}\,, \nonumber \\
({\cal M}^{(a)})^{--}_{++} & = & ({\cal M}^{(a)})^{++}_{--} \nonumber \\
   & = & \frac{i e^2}{t} (s+t) \,\bigg\{\frac{m_1^2}{2s} - i \frac{m_1^2}{s} 
   \sqrt{\frac{-t}{s+t}}\bigg\}\,, \nonumber \\
({\cal M}^{(a)})^{++}_{+-} & = & ({\cal M}^{(a)})^{++}_{-+} =
   ({\cal M}^{(a)})^{--}_{+-} = ({\cal M}^{(a)})^{--}_{-+} \nonumber \\
   & = & \frac{e^2}{2} \sqrt{\frac{s+t}{-t}} \,\bigg\{\frac{m_1}{\sqrt{s}}
   - i \frac{m_1^2}{s} \sqrt{\frac{-t}{s+t}}\bigg\}\,,
\label{calMajspap}
\end{eqnarray}
and all other components zero, where we have also neglected terms of the 
order of the electron mass $m$.  This approximation is already very good by 
$\sqrt{s}$ of order 10 GeV, at least near the forward direction where the
amplitudes are large, and becomes eventually necessary above this energy 
for computations without double precision. In Figure \ref{xsedexam} is 
shown the spin-summed differential cross sections for the reaction
$e^+ e^- \longrightarrow e^+ \tau^-$ at $\sqrt{s} =$ 10 GeV, 100 GeV
calculated with the rotation matrix element $S_{\alpha 1}$ of the
DSM scheme taken from ref. \cite{impromat,phenodsm}.  The curve at 10
GeV is calculated with the exact formulae (\ref{calMajsp}) which is
seen to be almost indistinguishable from the crosses calculated with
the approximate formulae (\ref{calMajspap}).  The curve at 100 GeV is
calculated with (\ref{calMajspap}) where the exact formulae is found
to have problems with accuracy in application.

Third, we note that the two sets of diagonal amplitudes (\ref{calMajsp}) 
and (\ref{calMbjsp}) were each calculated in a particular Lorentz frame,
namely for the diagram (a) in the cm frame of the $e^+ \ell_3^-$ system and
for the diagram (b) in the cm frame of the incoming $e^+ e^-$ system.
Although the amplitudes were all converted in the end into functions of the 
invariants $s$ and $t$, the directions of spin quantization are still
frame-dependent.  Hence, strictly speaking, the two frames for (a) and (b) 
being different, the two respective spin-amplitudes could not be added in 
the manner that we have done above.  However, the electron mass $m$ is so 
small compared to the energies we are interested in that this difference
in frame is entirely negligible for practical purposes.  Were the present
formalism to be adapted in future to say $\mu^+ \mu^-$ collisions, then
this would be a point to be borne in mind.  

With these points clarified, we have not encountered any more practical
difficulties in computing the cross sections of the three transmutation
reactions (\ref{etotau})--(\ref{eetomutau}).  Rather than presenting our
results in a wide range of $s$ and $t$, which could be confusing, we 
shall instead first give here a description of the general features, and 
then in the next section a detailed report on the result at $\sqrt{s}
= 10.58\ {\rm GeV}$, namely at the $\Upsilon(4S)$ where BaBar \cite{Babar},
Belle \cite{Belle}, and Cleo \cite{Cleo} have already collected a massive 
amount of data, in principle ready to be confronted with the above 
predictions.

\begin{figure}
\centering
\includegraphics[angle=-90,scale=0.55]{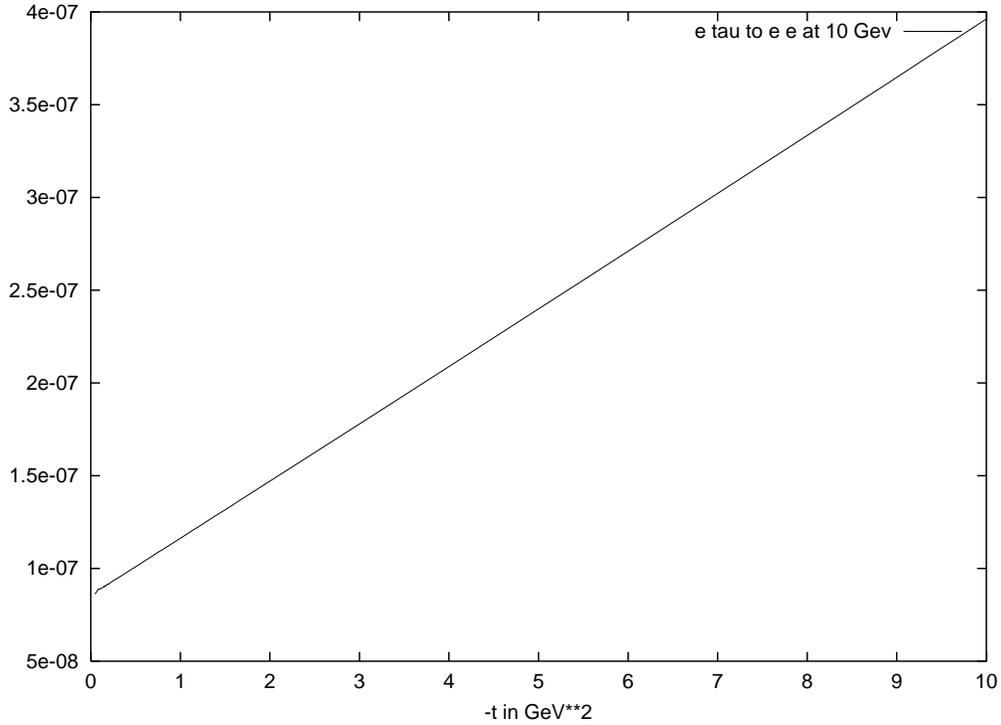}
\caption{The ratio at 10 GeV of the cross section of reaction $e^+ e^- 
\longrightarrow e^+ \tau^-$ over that of ordinary Bhabha scattering
$e^+ e^- \longrightarrow e^+ e^-$ as a function of $t$.}
\label{etautoee}
\end{figure}

Consider first the reactions (\ref{etotau}) and (\ref{etomu}) which receive 
contributions from both diagrams (a) and (b), and are very similar except 
for the difference in normalization due to the different values of the 
rotation matrix elements $S_{\alpha 1}$.  As in ordinary Bhabha scattering, 
the amplitude (a) is dominated by the pole at $t=0$ which gives the cross 
section a sharp forward peak, as can be seen in the examples of Figure 
\ref{xsedexam}.  Except at large scattering angles where $t$ is of order $s$, 
this peak overshadows the contribution from the (b) diagram.  However, the 
forward peak for the transmutation reactions (\ref{etotau}) and (\ref{etomu}) 
is nowhere near as sharp as for ordinary Bhabha scattering, as can be 
seen in Figure \ref{etautoee}.  The reason for this difference is seen 
in (\ref{calMajspap}), where one notices that the normally dominant spin 
non-flip amplitudes with $1/t$ behaviour are of order $m_1^2/s$, while 
the spin flip with a weaker $1/\sqrt{-t}$ behaviour are of order 
$m_1/\sqrt{s}$.  The same formulae (\ref{calMajspap}) explains also the
sharp decline of the cross section with increasing energy as well as 
its change in $t$-dependence as the spin flip terms become ever more
dominant, both of which effects can be seen in Figure \ref{xsedexam} by
comparing the curves at $\sqrt{s}=10$ and $100$ GeV.

\begin{figure}
\centering
\includegraphics[angle=-90,scale=0.55]{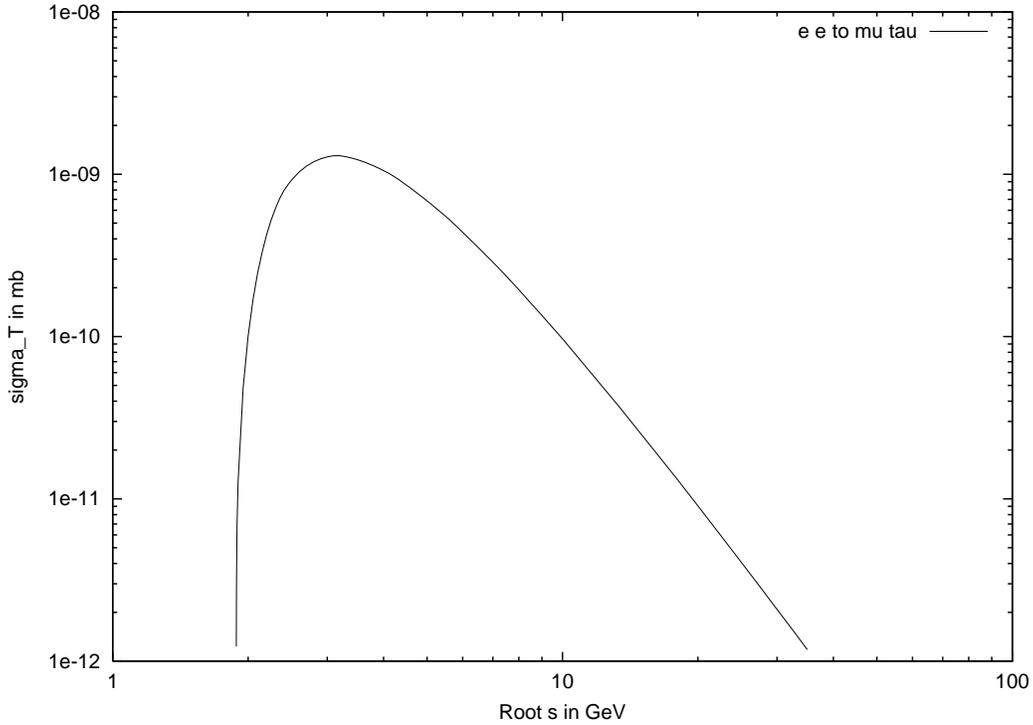}
\caption{Cross section for the reaction $e^+ e^- \longrightarrow \mu^+ 
\tau^-$ integrated over all scattering angles as a function of 
$\sqrt{s}$.}
\label{xsectot}
\end{figure}

The other reaction (\ref{eetomutau}) receives contributions only from
the (b) diagram which has no peak in the forward direction.  It is 
distinguished from the same diagram in ordinary Bhabha scattering by the
fact that, like the (a) transmutation amplitude, it is also dominated by
the spin flip terms at high energy.  Without the sharp singular peak in the
forward direction, it gives, in contrast to reactions (\ref{etotau}) and 
(\ref{etomu}), a finite total cross section, the rough energy dependence of 
which is shown in Figure \ref{xsectot}, where one sees that, as in
photo-transmutation \cite{photrans}, the cross section rises shortly
after threshold to a peak and then declines as $\sqrt{s}$ increases. 

We note that since the rotation matrix elements $S_{\alpha j}$ enter only 
in the normalization of the cross section and are themselves only weakly 
dependent on energy, their actual values do not affect the discussion above 
on the the $t$-dependence, and qualitatively also on the $s$-dependence, 
of the cross section.

\section{Transmutation Cross Sections at $\sqrt{s} = 10.58$ GeV}

The reason for selecting this particular energy corresponding to the mass
of the $\Upsilon(4S)$ for detailed analysis is that two experiments of
ultra-high sensitivity, namely BaBar and Belle, are running and have
collected already up to 20 ${\rm fb}^{-1}$ of luminosity, with much more
expected in the near future \cite{Babar,Belle}.  Although these experiments 
were designed originally to look for other rare effects like CP-violation 
from $B$ decay, their data could conveniently be used also to search for 
the transmutation reactions of interest to us here. 

\begin{figure} 
\centerline{\psfig{figure=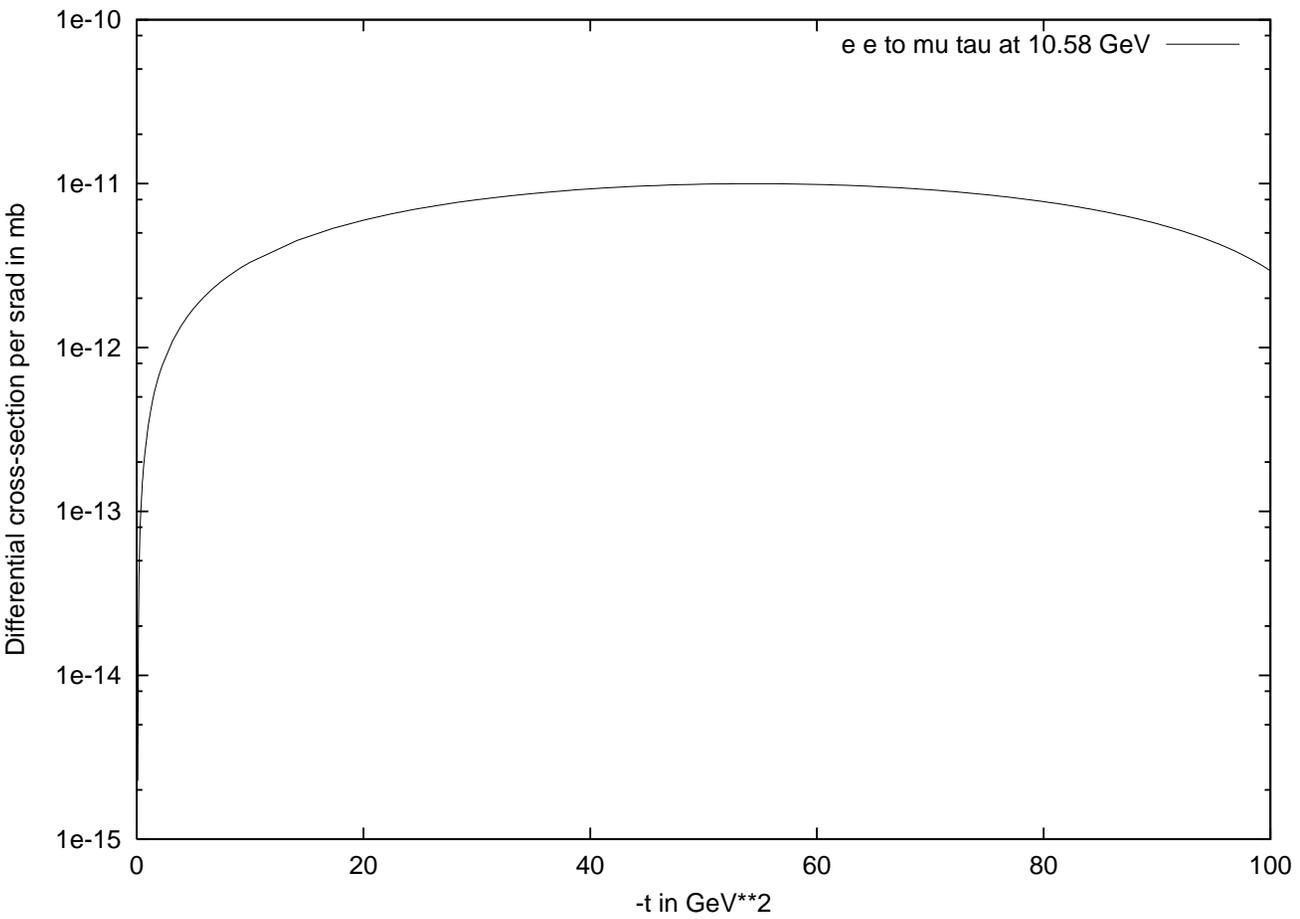,width=0.9\textwidth}}
\caption{Spin-summed differential cross section for the reaction $e^+ e^-
\longrightarrow \mu^+ \tau^-$ at $\sqrt{s} = 10.58\ {\rm GeV}$. For the
normalization, see text.}
\label{dxsecmutau}
\end{figure}

Consider first the reaction (\ref{eetomutau}) with contributions only 
from Figure \ref{Feyndi}(b).  The normalization of the cross section is,
according to (\ref{calMrewrite}), given by the rotation matrix elements
$S_{\mu 1} S_{\tau 1}$.  If one accepts the contention of \cite{empirdsm}
that it is the rotation which is giving rise to fermion mixing and mass 
hierarchy, then these rotation elements should be related to the data
on the mixing parameters and mass ratios.  Indeed, as explained in that
paper, the quantity $S_{\mu 1} S_{\tau 1}$ would then be given to a good
approximation by $\sin \theta \cos \theta$ where $\theta$ is the rotation
angle from the $\tau$-mass scale where the mass matrix is diagonal in the
lepton flavour states to the scale 10.58 GeV of say the BaBar experiment.
Its value can thus be read off from Figure 3 of \cite{empirdsm}, either
directly by interpolating the actual data points or equivalently, since
the rotation curve from the DSM scheme is seen there to be a very good
fit to the data, by taking the values from the DSM calculation, in either
case leading to an estimate $S_{\mu 1} S_{\tau 1} \sim 0.043$.  With 
this then for the normalization, one obtains Figure \ref{dxsecmutau} for 
the spin-summed diffrential cross section of reaction (\ref{eetomutau}).
Integrating over the whole angular range, one obtains a cross section
of around 80 fb.  This can also be read in Figure \ref{xsectot} which was 
in fact calculated already with this normalization.  In practical terms, 
this could mean as many as 1600 events in the data sample of 20 fb$^{-1}$
already collected by BaBar, assuming 100 percent efficiency.  

\begin{figure}
\centerline{\psfig{figure=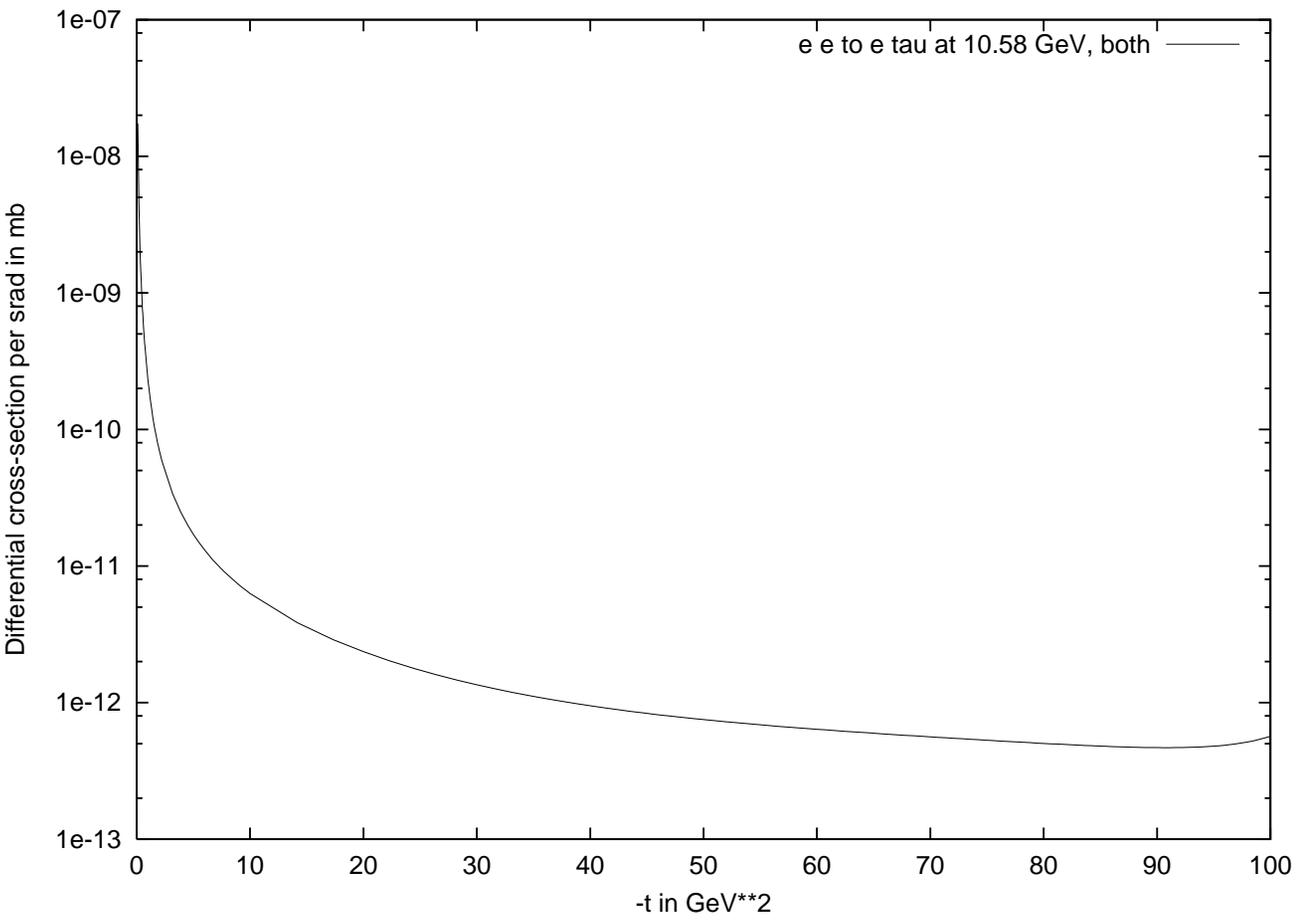,width=0.9\textwidth}}
\caption{Spin-summed differential cross section for the reaction $e^+ e^-
\longrightarrow e^+ \tau^-$ at $\sqrt{s} = 10.58\ {\rm GeV}$.  For the
normalization see text.}
\label{xsec}
\end{figure}

Consider next the reaction (\ref{etotau}) which receives contribution 
from both diagrams in Figure \ref{Feyndi}.  Its normalization depends
now instead on the rotation matrix elements $S_{e 1} S_{\tau 1}$ which
is considerably smaller and harder to estimate directly from data.  Let 
us then just insert the values obtained from the DSM scheme \cite{phenodsm}
which was shown \cite{empirdsm} to fit the data very well and can be
taken as a convenient interpolation device.  The actual value can be
read in Figure 4 of \cite{photrans} at 10.58 GeV to be around 0.0092.
This then gives Figure \ref{xsec} for the spin-summed differential cross 
section for this reaction.  As in ordinary Bhabha scattering, the cross 
section for (\ref{etotau}) is divergent at $t=0$.  However, this region 
cannot be explored experimentally, the detectors being insensitive to 
the forward region with $|t| \lesssim 5\ {\rm GeV}^2$ \cite{Babar}.  A 
rough estimate from Figure \ref{xsec} then yields for the integrated cross 
section for reaction (\ref{etotau}) over the range $|t| < 5\ {\rm GeV}^2$ 
a value of about 20 fb, which means about 400 events in the sample of 20
fb$^{-1}$ already collected in last year's run assuming again 100 percent
detection efficiency.  Compared to reaction (\ref{eetomutau}) above, we 
note the very different angular dependence which is here dominated by the 
diagram (a) of Figure \ref{Feyndi} absent in reaction (\ref{eetomutau}).
The reason why the cross section for (\ref{etotau}) is smaller than that 
for (\ref{eetomutau}) despite the fact that (\ref{eetomutau}) receives 
contribution only from the sub-dominant diagram (b) is the very different
sizes of the rotation matrix elements involved, $(S_{\mu 1}/S_{e1})^2$
being $\sim 20$, as read from Figure 4 in \cite{photrans}.   

However, we should perhaps stress once more that the Figures \ref{dxsecmutau} 
and \ref{xsec} calculated by the method of \cite{photrans} represent only the
kinematic effects of the rotating mass matrix.  In particular, though 
calculated with DSM rotation matrices, the transmutation cross sections 
given are not those predicted by the DSM scheme in which, as seen in 
\cite{transmudsm}, the above kinematic effects are largely cancelled by 
other rotation effects implied concurrently by the renormalization mechanism 
driving the mass matrix rotation. 

Apart from minor kinematic differences, the spin-summed differential cross 
section for the reaction (\ref{etomu}) has the same $t$-dependence as
(\ref{etotau}) but a different normalization, namely with $S_{\tau 1}$ in 
(\ref{etotau}) replaced by $S_{\mu 1}$ in (\ref{etomu}).  For instance, 
taking again the rotation matrix elements from Figure 4 of \cite{photrans}, 
one obtains a cross section for (\ref{etomu}) a factor $\sim 2 \times 10^{-3}$
smaller than for (\ref{etotau}), making it probably difficult in any case
to observe 
in the near future.

For all 3 reactions, there is in principle also a contribution from the 
transmutational decay of the $\Upsilon (4S)$ resonance with this mass,
but this will be seen in the following section to be negligible in
comparison with the above contributions.  

From the above results, it would appear that if one accepts the interpretation
as given in \cite{empirdsm} of fermion mixing and mass hierarchy as rotation
effects, and that there are no other rotation effects than that of the mass
matrix, then there will be lepton flavour-violating transmutation effects 
in $e^+ e^-$ collisions at BaBar, Belle, and Cleo energy, which are of a
magnitude to be observable by these experiments.  On the other hand, if the
interpretation of \cite{empirdsm} is not accepted, one knows at present of no 
other empirical means for estimating the rotation angles and thus no estimate 
for the absolute rates of transmutation can yet be made.  However, since the
rotation matrix elements enter only in the normalization of the cross section,
the calculation above is easily adaptable to any rotation matrix obtained
from whatever source, whether empirical or theoretical.

Suppose that the reactions (\ref{eetomutau}) and 
(\ref{etotau}) are indeed observed, can one be sure that they are due to 
transmutation and not some other lepton-violating effect?  The answer to 
this question would seem to be quite affirmative since one has here the
differential cross sections as functions of 2 variables, each with
distinctive characteristics.   For example, it is predicted that the
cross section for (\ref{etotau}) should be peaked sharply forwards
as seen in Figure \ref{xsec}, while for (\ref{eetomutau}) it should have 
a roughly $\sin^2 \theta$ behaviour, as seen in Figure \ref{dxsecmutau}.  
And both are predicted to be spin-flip dominated, which assertion may 
be verifiable with the decaying $\tau$ serving as its own spin-analyser 
and the spin-dependent cross section calculable from the amplitudes 
(\ref{calMajsp}) and (\ref{calMbjsp}) when the occasion demands.  
Thus, if the reactions are observed at all with any reasonable statistics,
the signatures for transmutation would seem to be quite unmistakable.

We have restricted the discussion in this section to only the operation 
energy of BaBar, Belle and Cleo, but very similar remarks apply also to
the BEPC energy \cite{Bepc}.  The expected transmutation cross sections 
at BEPC, as seen in Figure \ref{xsectot} for reaction (\ref{eetomutau}), 
are even larger, but this advantage is unfortunately more than offset by 
the lower luminosity so far achieved by the machine.  For this reason, 
for BEPC, we think a search for transmutation in $\psi$ decay will be more  
immediately profitable, as we shall elucidate in the following section.  
As for LEP, according to Figure \ref{xsectot} for example, transmutation 
cross sections would have fallen much below the fb level by that energy 
even for the optimistic scenario of \cite{empirdsm} and are thus unlikely 
to show up in the data collected.

\section{Vector Boson Decay}

Vector boson formation and decay occur naturally in $e^+ e^-$ collision
so that they will have in any case to be taken into account in studying 
transmutation from this process.  Besides, being essentially a single
particle effect, vector boson decays are easier to analyse theoretically 
and can give more succint conclusions than the reactions previously 
considered.  We propose therefore to examine in this section the following
transmutation decays of vector bosons into leptonic final states:  
\begin{equation}
V \longrightarrow \ell_\alpha \bar{\ell}_\beta,  
\label{Vtollbar}
\end{equation}
for $\ell_\alpha (\ell_\beta)$ being $e, \mu$ or $\tau$.  Order of magnitude 
estimates for some of these decays have already been made in \cite{impromat}.  Now, with the procedure developed in \cite{photrans} an explicit calculation 
can be carried out.

Consider then in general a decay of the type (\ref{Vtollbar}) as depicted 
in Figure \ref{decaydiag}(a).  The decay amplitude is to be evaluated at 
the scale $\sqrt{s} = M$, with $M$ being the mass of the decaying boson.  
At this scale, the fermion mass matrix $m$ is in general not diagonal in 
the flavour states $e, \mu, \tau$ but in some other states, say, $i = 1,2,3$ 
with masses (eigenvalues of $m$) $m_i$, the two triads of state vectors 
being related by a rotation matrix $S_{\alpha i} = \langle \alpha|i \rangle$.  
To calculate the decay amplitude, we first evaluate the amplitudes for the 
decays into the diagonal states $i$, namely $V \longrightarrow \ell_i 
\bar{\ell}_i$
as depicted in Figure \ref{decaydiag}(b).  For a vector boson with polarization
vector $\epsilon_a^\mu$, this amplitude is:
\begin{equation}
{\cal M}_i = g \epsilon_a^\mu(k) \bar{u}_r(p_i) \gamma_\mu v_s(q_i),
\label{calMi}
\end{equation}
where $r$ and $s$ denote the spin respectively of $\ell_i$ and 
$\bar{\ell}_i$.  We
do not need to specify the coupling strength $g$ of the vector boson to 
the lepton pair for in our branching ratios calculation it will be cancelled 
out.  According to the procedure suggested in \cite{photrans}, the amplitude 
for the decay of interest (\ref{Vtollbar}) is then given just by a rotation 
of these diagonal amplitudes ${\cal M}_i$ to the appropriate lepton flavour 
states, thus:
\begin{equation}
{\cal M}^{\alpha \beta} = \sum_i S^{\dagger}_{\alpha i} {\cal M}_i 
S_{\beta i}.
\label{calMab}
\end{equation}

To obtain the decay rate, with which alone we shall be concerned at present,
we average over the initial polarization $a$ of the decaying boson and sum
over the final spins $r, s$ of the product leptons, obtaining:
\begin{equation}
\frac{1}{3} \sum_{a,r,s} |{\cal M}^{\alpha \beta}_{ars}|^2 
   = -\frac{1}{3} g^{\mu \nu} \sum_{r,s} {\cal M}^{\alpha \beta}_{\mu rs} 
      ({\cal M}^{\alpha \beta}_{\nu rs})^{\dagger}
   + \frac{1}{3} \frac{1}{M^2} \sum_{r,s} |{\cal M}^{\alpha \beta}_{(k)rs}|^2,
\label{sumars}
\end{equation}
where
\begin{eqnarray}
{\cal M}^{\alpha \beta}_{\mu rs} & = & g \sum_i S^{\dagger}_{\alpha i} 
   \bar{u}_r(p_i) \gamma_\mu v_s(q_i) S_{\beta i}, \\
{\cal M}^{\alpha \beta}_{(k) rs} & = & g \sum_i S^{\dagger}_{\alpha i} 
   \bar{u}_r(p_i) k\llap/ v_s(q_i) S_{\beta i}.
\label{Mmuk}
\end{eqnarray}
The remaining sum over lepton spins again
cannot readily be done as an invariant 
trace because of the crossed terms between different internal channels $i$.  
We proceed as for the reactions (\ref{etotau})---(\ref{eetomutau}) above. 

We choose to work in the rest frame of the decaying vector boson:
\begin{equation}
{\bf k} = {\bf p}_i + {\bf q}_i = 0,
\label{restframe}
\end{equation}
so that:
\begin{eqnarray}
p_i^\mu & = & (E, 0, 0, -\omega_i), \nonumber \\
q_i^\mu & = & (E, 0, 0, \omega_i), \nonumber \\
k^\mu & = & (M, 0, 0, 0),
\label{piqik}
\end{eqnarray}
with 
\begin{equation}
E = M/2, \ \ \ \omega_i = \sqrt{E^2 - m_i^2}.
\label{Eomegai}
\end{equation}
As a result, we have
\begin{eqnarray}
u_+(p_i)  =  \frac{1}{\sqrt{2(E+m_i)}} \left(\!\! \begin{array}{c} 
   0 \\ E+m_i \\ 0 \\ \omega_i  \end{array} \!\!\right)\!\!\!\!\!&;&  
\!\!\!\!u_-(p_i) =  \frac{1}{\sqrt{2(E+m_i)}} \left(\! \begin{array}{c} 
   E+m_i\\0  \\ -\omega_i \\ 0 \end{array}\! \right); \nonumber \\
v_+(q_i)  =  \frac{1}{\sqrt{2(E+m_i)}} \left(\!\! \begin{array}{c} 
   0 \\ -\omega_i \\ 0 \\ E+m_i \end{array} \!\right)\!\!\!\!\!&;&  
\!\!\!\!v_-(q_i) =  \frac{1}{\sqrt{2(E+m_i)}} \left(\! \begin{array}{c} 
   \omega_i \\ 0 \\ E+m_i\\ 0  \end{array} \!\!\right)
\label{uv}
\end{eqnarray}
With these explicit expressions for the lepton wave functions, it is easy
to evaluate the spin amplitudes and perform the sum over the lepton spins
giving:
\begin{equation}
\frac{1}{3} \sum_{a,r,s} |{\cal M}^{\alpha \beta}_{ars}|^2 
   = \frac{4 g^2}{3} \delta_{\alpha \beta} E^2 + \frac{2 g^2}{3} \sum_{i,j}
     S_{\alpha i}^{\dagger} S_{\beta i} m_i m_j S_{\beta j}^{\dagger} 
     S_{\alpha j}.
\label{sumarsa}
\end{equation}
The states $i$ being by definition the eigenstates of the mass matrix $m$
at the scale of the decaying boson mass $M$, and $S_{\alpha i}$ the rotation
matrix relating these states $i$ to the lepton flavour state $\alpha = e, \mu$,
or $\tau$, the sum $\sum_i S_{\alpha i}^{\dagger} S_{\beta i} m_i$ is just
the element $\langle \alpha|m|\beta \rangle$ of the matrix $m$ at the boson 
mass scale.  Hence,
\begin{equation}
\frac{1}{3} \sum_{a,r,s} |{\cal M}^{\alpha \beta}_{ars}|^2 
   = \frac{4 g^2}{3} \delta_{\alpha \beta} E^2 
     + \frac{2 g^2}{3} |\langle \alpha|m|\beta \rangle|^2.
\label{sumarsb}
\end{equation}

This gives the total width for the decay $V \longrightarrow \ell_\alpha 
\bar{\ell}_\beta$ as:
\begin{equation}
\Gamma = \frac{g^2}{\pi}\frac{1}{4M} \frac{\omega_{\alpha \beta}^2}
   {E_\alpha E_\beta} \frac{1}{3} \left[M^2 \delta_{\alpha \beta}
   + 2 |\langle \alpha|m|\beta \rangle|^2 \right],
\label{Gammat}
\end{equation}
where we note that the phase space factor $\omega_{\alpha \beta}^2/
E_\alpha E_\beta$ refers as per \cite{photrans} to the freely propagating
``external'' leptons, with momentum and energy respectively:
\begin{eqnarray}
\omega_{\alpha \beta} & = & \frac{1}{2M} \sqrt{M^4 + m_\alpha^4 + m_\beta^4
   - 2M^2 m_\alpha^2 - 2 M^2 m_\beta^2 - 2 m_\alpha^2 m_\beta^2}, \nonumber \\
E_\alpha & = & \frac{1}{2M}(M^2 + m_\alpha^2 - m_\beta^2), \nonumber \\
E_\beta & = & \frac{1}{2M}(M^2 - m_\alpha^2 + m_\beta^2).
\label{omegaEab}
\end{eqnarray}
More conveniently, since the coupling $g$ has not been specified, one writes
for $\alpha \neq \beta$, i.e. transmutational decays:
\begin{equation}
\frac{\Gamma(V \longrightarrow \ell_\alpha \bar{\ell}_\beta)}
   {\Gamma(V \longrightarrow e^+ e^-)} = \frac{\omega_{\alpha \beta}^2}
   {E_\alpha E_\beta} \frac{2}{M^2} |\langle \alpha|m|\beta \rangle|^2,
\label{BRe}
\end{equation}
where we have neglected terms of the order of the electron mass $m_e$ 
compared to $M$.  Multiplying then this ratio by the experimental branching 
ratio, if known, of the boson $V$ decaying into $e^+ e^-$ gives the 
branching ratio of the transmutational $\ell_\alpha \bar{\ell}_\beta$ mode.

One notices that apart from a numerical factor of $2$ and the phase space 
factor $\omega_{\alpha \beta}^2/E_\alpha E_\beta$, the formula for the 
branching ratio (\ref{BRe}) for transmutational decays is the same as the 
order-of-magnitude estimate given in \cite{impromat}.  This formula is 
supposed to hold in general and can be applied to calculate the branching 
ratio of transmutational decays of the type (\ref{Vtollbar}) given the 
matrix element $\langle \alpha|m| \beta \rangle$ at scale $M$.

For a numerical example, we take again the interpretation in \cite{empirdsm}
of experimental data to estimate the rotation matrix elements required or,
in case the data is insufficient, employ the DSM result of \cite{phenodsm}
as an interpolation formula, for which the required mass matrix elements
$\langle \alpha|m| \beta \rangle$ can be read directly from Figure 3 of
\cite{impromat}.  With these elements, the formula (\ref{BRe}) can be applied
immediately to calculate the branching ratios of transmutational modes
in the decay of any vector boson by normalizing to the empirical branching 
ratios of the $e^+ e^-$ mode (or to the $\mu^+ \mu^-$ mode if more accurate) 
given in \cite{databook}.  The result for the most experimentally interesting 
vector bosons listed in \cite{databook} is given in Table \ref{BRtrans}.
(Again, the reader is reminded that these numbers represent only the
transmutation effects due to kinematics of the rotating mass matrix alone 
but not, though calculated with DSM rotation matrices, predictions of the 
DSM scheme in which cancellations with other rotational effects occur 
\cite{transmudsm}.)  We have calculated also the branching ratios for 
the other vector bosons listed in \cite{databook}, such as the higher 
excitations of $\psi$, but for these, once the mass gets above the 
$D \bar{D}$ threshold, hadronic decays prevail, leading to large total 
widths and hence uninterestingly small branching ratios for the modes 
that concern us here, as can be seen in the example given of $\psi(3770)$.  
In any case, the predicted branching ratios for all higher excitations of 
$\psi$ and $\Upsilon$ are very similar to those given in Table \ref{BRtrans} 
for $\psi(3770)$ and $\Upsilon(10860)$ and are therefore not given again 
there.

\begin{table}
\begin{eqnarray*}
\begin{array}{||c||c||c||}  
\hline \hline
{\rm Boson} & {\rm Mode} & {\rm Predicted\ Branching\ Ratio} \\         
\hline \hline
\phi(1020) & e \mu & 2.5 \times 10^{-12} \\ \hline
\psi(1S) & \mu \tau & \underline{6.3 \times 10^{-6}} \\ 
     & e \tau   & 1.7 \times 10^{-7} \\ 
     & e \mu    & 1.1 \times 10^{-10} \\ \hline
\psi(2S) & \mu \tau & \underline{1.2 \times 10^{-6}} \\ 
         & e \tau   & 3.8 \times 10^{-8} \\ 
         & e \mu    & 3.1 \times 10^{-11} \\ \hline
\psi(3770) & \mu \tau & 1.7 \times 10^{-9} \\ 
           & e \tau   & 5.3 \times 10^{-11} \\ 
           & e \mu    & 4.3 \times 10^{-14} \\ \hline 
\Upsilon(1S) & \mu \tau & \underline{2.9 \times 10^{-6}}\\ 
           & e \tau   & \underline{1.4 \times 10^{-7}} \\ 
           & e \mu    & 2.6 \times 10^{-10} \\ \hline
\Upsilon(2S) & \mu \tau & \underline{1.3 \times 10^{-6}} \\ 
             & e \tau   & 6.2 \times 10^{-8} \\ 
             & e \mu    & 1.2 \times 10^{-10} \\ \hline
\Upsilon(3S) & \mu \tau & \underline{1.9 \times 10^{-7}} \\
             & e \tau   & 9.3 \times 10^{-9} \\
             & e \mu    & 1.8 \times 10^{-11} \\ \hline
\Upsilon(4S) & \mu \tau & 2.8 \times 10^{-9} \\
             & e \tau   & 1.4 \times 10^{-10} \\
             & e \mu    & 2.8 \times 10^{-13} \\ \hline
\Upsilon(10860) & \mu \tau & 2.7 \times 10^{-10} \\
                & e \tau   & 1.4 \times 10^{-11}\\
                & e \mu    & 2.8 \times 10^{-14} \\ \hline
Z_0 & \mu \tau & \underline{1.0 \times 10^{-7}} \\ 
    & e \tau   & 8.8 \times 10^{-9} \\ 
    & e \mu    & 3.5 \times 10^{-11} \\ \hline \hline
\end{array}
\end{eqnarray*}
\caption{Predicted branching ratios for transmutational decays.  Modes 
which could be accessible to experiment in the near future are underlined.}
\label{BRtrans}
\end{table} 

In Table \ref{BRtrans}, we note first that the predicted branching ratios 
for flavour-violating transmutational decays even for the fast rotation 
implied by \cite{empirdsm} are not as large as one might fear at first 
sight.  Indeed, all the estimates survive existing experimental bounds 
comfortably, which are surprisingly weak for vector boson decays.  Of the 
vector bosons listed, the data book \cite{databook} gives upper limits 
on lepton-flavour violations only for $Z_0$ decay, which for the modes 
$\mu \tau, e \tau$, and $e \mu$ are respectively $1.2 \times 10^{-5}, 
9.8 \times 10^{-6}$, and $1.7 \times 10^{-6}$ in branching ratios, which 
are seen to be all easily satisfied by the estimates in Table \ref{BRtrans}.

Secondly, we note several entries of branching ratios in Table \ref{BRtrans} 
fall well within the sensitivity range of present experimental set-ups.  In 
particular, the $\mu \tau$ decay modes of $\psi(1S), \psi(2S), \Upsilon(1S)$, 
and $\Upsilon(2S)$, are seen each to have a predicted branching ratio of 
several parts in a million and hence well within the present sensitivity 
range of BEPC \cite{Bepc}, PEP II (BaBar) \cite{Babar} and BELLE \cite{Belle}.
For instance, BEPC has already collected more than 20 million $\psi$'s 
to-date, and expects to collect twice as many more next year, which would 
mean for the predicted branching ratio of $6.3 \times 10^{-6}$ as many as 
120 now, and 360 next year, of $\mu \tau$ decays assuming 100 percent 
detection efficiency.  Similarly, PEP II has already accumulated for BaBar 
in a year over 20  ${\rm fb}^{-1}$ in luminosity at $\Upsilon(4S)$, which 
means that if the machine is run at the $\Upsilon(1S)$ with a cross section 
of around 25 nb, it would collect in just a couple of months of running 
already enough $\Upsilon(1S)$ events to give, at a branching ratio of 
$1.2 \times 10^{-6}$, over 100 $\mu \tau$ decays, again assuming 100 
percent detection efficiency.  We note in passing, however, that at the
energy 10.58 GeV at which the present BaBar experiment is run, the
$\Upsilon(4S)$ resonance at that mass, being above the $B \bar{B}$ hadron
threshold, has a branching ratio of only $2.8 \times 10^{-9}$ into the
already most copious transmutational $\mu \tau$ mode, which makes the
resonance effect negligible compared with direct transmutation via the
process of Figure \ref{Feyndi}(b) as calculated in the preceding section.

\section{Concluding Remarks}

In summary, our conclusions are as follows.  Although flavour conservation
has been checked to very high accuracy in leptonic decays such as 
$\mu \longrightarrow e \gamma$ and $\mu \longrightarrow e e \bar{e}$,
it may be premature to conclude that flavour-violation will always be
small.  The fact that the fermion mass matrix can rotate with changing
scale can mean that flavour though accurately conserved at some energies
are appreciably violated at other energies.  We suggest therefore that
lepton flavour violation should be routinely tested by experiment whenever
conditions are favourable.  Any sizeable signal (e.g.\ BR in vector boson
of say order greater than $10^{-13}$ \cite{Nussinov}) can mean mass matrix 
rotation, and any information on rotation, and especially on its energy 
dependence, can give us new insight into the problem of fermion generations.

If we make the assumption that fermion mixing and mass hierarchy are due
to mass matrix rotation, for which hypothesis there seems to be some
empirical support \cite{empirdsm}, then the magnitude of the kinematic
effects on flavour-violating from a rotating mass matrix can be estimated, 
and have been found to be appreciable and very likely detectable with the 
present generation of high sensitivity experiments such as BaBar, Belle, 
Bepc and Cleo.  A detection of the effect at the estimated level would 
lend support to the hypothesis of \cite{empirdsm} which would be a big 
step forward towards the solution of the generation puzzle.  However, a 
negative result, unfortunately, cannot rule out the rotation hypothesis, 
for it can happen, as it is shown to be actually the case in a specific
example in \cite{transmudsm}, that the effect of the rotating mass matrix 
calculated here is modified or even cancelled by other rotation effects 
accompanying the mass matrix rotation.  

We thank Gian Gopal and John Guy for useful discussions on the experimental
feasibility of testing the ideas of this paper and for their encouragement
through all the ups and downs of its progress.

\end{document}